\documentclass[aps,twocolumn,superscriptaddress,prl,floatfix]{revtex4}
\usepackage{amsmath}
\usepackage{amssymb}

\usepackage{siunitx}
\usepackage{graphicx}
\usepackage{dcolumn}
\usepackage{bm}
\usepackage{xfrac}

\usepackage{xcolor}
\usepackage{float}

\usepackage{hyperref}
\hypersetup{
    colorlinks=true,
    linkcolor=blue,
    filecolor=magenta,      
    urlcolor=cyan,
}

\begin{document}

\title{Anomalous Phase Dynamics of Driven Graphene Josephson Junctions}
\author
{S. S. Kalantre,$^{1 \dag}$ F. Yu,$^{1 \dag}$ M. T. Wei$^{1}$, K. Watanabe$^{2}$, T. Taniguchi$^{2}$, M. Hernandez-Rivera$^{3}$, F. Amet$^{3}$, and J. R. Williams$^{1 \ast}$
\\
\vspace*{0.25cm}
\small{$^{1}$\emph{Joint Quantum Institute and Center for Nanophysics and Advanced Materials}}\\
\small{\emph{Department of Physics, University of Maryland, College Park, MD USA.}}\\
\small{$^{2}$\emph{Advanced Materials Laboratory, National Institute for Materials Science}}\\
\small{\emph{1-1 Namiki, Tsukuba, 305-0044, Japan.}}\\
\small{$^{3}$\emph{Department of Physics and Astronomy, Appalachian State University, Boone, NC, USA}}\\
\footnotesize{$^\dag$\emph{These authors equally contributed.}}\\
\footnotesize{$^\ast$\emph{To whom correspondence should be addressed; E-mail:  jwilliams@physics.umd.edu.}}\\
}

\date{\today}

\begin{abstract}
Josephson junctions with weak-links of exotic materials allow the elucidation of the Josephson effect in previously unexplored regimes. Further, such devices offer a direct probe of novel material properties, for example in the search for Majorana fermions. In this work, we report on DC and AC Josephson effect of high-mobility, hexagonal boron nitride (h-BN) encapsulated graphene Josephson junctions. On the application of RF radiation, we measure phase-locked Shapiro steps. An unexpected bistability between $\pm 1$ steps is observed with switching times on the order of seconds. A critical scaling of a bistable state is measured directly from the switching time, allowing for direct comparison to numerical simulations. We show such intermittent chaotic behavior is a consequence of the nonlinear dynamics of the junction and has a sensitive dependence on the current-phase relation. This work draws connections between nonlinear phenomena in dynamical systems and their implications for ongoing condensed matter experiments exploring topology and exotic physics.

\end{abstract}

\maketitle

The ground state wavefunction of a superconductor is endowed with an emergent phase on the macroscopic scale~\cite{tinkham2004introduction}. Josephson pointed out that two superconductors separated by a non-superconducting weak link produces a supercurrent that depends on the phase difference between the superconductors~\cite{josephson_possible_1962}. Today, eponymously known as the Josephson effect, it remains a hallmark of the superconducting state and has led to a large number of scientific and technological advances~\cite{LikharevJJ, PaternoBarone, devoret_superconducting_2013}. As one of the few phase-sensitive devices available in condensed matter physics, it has found use in the investigation of novel superconducting materials~\cite{vanHarlingen95} and more recently in the identification of topological superconductors and Majorana bound states~\cite{rokhinson_fractional_2012, wiedenmann_4_2016, laroche_observation_2019, Wang2018, Yu2018, Calvez19}. In particular, the Josephson effect in the presence of RF radiation (the AC Josephson effect) has been instrumental in advancing our knowledge of underlying physics in these systems. The incorporation of high-quality, tunable two-dimensional materials into a Josephson junction (JJ) now allow for these nonlinear devices to be probed in previous unexplored regimes. As the JJ becomes more prevalent in use both as a detector of novel materials properties and exotic circuit elements, a thorough understanding of the role of nonlinear dynamics in these devices is essential. 

The understanding of JJs driven by externally applied currents and radio frequency (RF) radiation rests on simple circuit models that capture the superconducting phase dynamics. The relationship between the supercurrent ($I$) and the phase difference across the JJ ($\phi$) is encoded in a current-phase relationship (CPR), which subsumes the microscopic physics of the weak-link \cite{golubov_current-phase_2004}. The dynamics of a JJ then follow the evolution of a phase particle in an effective potential dependent on the physical parameters of the junction, such as normal state resistance, capacitance and externally applied currents \cite{stewart_currentvoltage_1968,mccumber_effect_1968}. At the heart of such models is a dynamical system described by a set ordinary differential equations made nonlinear by the CPR \cite{tinkham2004introduction, titov_josephson_2006}. 

Nonlinear dynamical systems are interesting in their own right because of the rich set of phenomena they display. The phase dynamics of a driven JJ is similar to that of a driven, damped pendulum -- a system known to be chaotic. Nonlinear effects in JJs have been studied extensively in theory \cite{huberman_noise_1980, kautz_ac_1981, kautz_chaotic_1981, pedersen_chaos_1981, cirillo_bifurcations_1982, kautz_chaos_1985, kautz_survey_1985}. As a result, a zoo of nonlinear effects including chaos, intermittency and strange attractors can occur \cite{ott2002chaos}, and these effects should be manifest in JJs. Hence, it is critical to detail these phenomena to have a thorough understanding of the physics of JJs and to ensure these effects do not masquerade as any exotic effects that could erroneously be ascribed to the novelty of the weak link material.

In this paper, we draw a bridge between experiments on graphene JJs and the unexpected consequences of the nonlinear phase dynamics. Graphene encapsulated in hexagonal boron nitride (h-BN) offers a platform to fabricate junctions with ballistic carrier transport. Moreover, gate electrodes and magnetic fields can be used to tune the junction parameters. We focus on measurements of the junction resistance as a function of a DC current bias and RF power, referred to as Shapiro diagrams. Measurements the Shapiro diagram show strong deviations from the expected behavior. In particular, portions of these diagram disappear in certain regimes, a result of chaotic behavior possessing timescales many orders of magnitude slower than have ever been observed before in JJs. This slow time scale allows for this nonlinear effect to be tracked in real time. Through comparison to simulations, the time scale is shown to depend sensitively on the CPR of the junction, becoming large for skewed CPRs like that predicted for weak links materials possessing a Dirac (linear) energy dispersion. These results should inform future experiments using RF radiation to probe exotic materials in JJs~\cite{rokhinson_fractional_2012, wiedenmann_4_2016, laroche_observation_2019} and devices that exploit JJs in RF environments~\cite{gTrans1, gTrans2}.   

\section{Device Characteristics}
h-BN encapsulated graphene is fabricated into JJs, proximitized with molybdenum-rhenium alloy (MoRe) superconducting leads [Fig.~\ref{fig1}(a)]. MoRe is a type-II superconductor with an upper critical field of 8 \si{\tesla}. Previously, such junctions have been used to study supercurrent in the quantum Hall regime and have been shown to have ballistic carrier transport \cite{amet_supercurrent_2016}. The junction length along the direction of current flow is  $\sim$500 \si{\nano\meter} and the width is $\sim$2.7 \si{\micro\meter}. In addition to MoRe contacts, we also have Cr/Au top gates and a backgate to electrically tune electron density. The top gates are kept grounded for the entirety of this experiment. 

The device was cooled down in a cryostat with a base temperature of $50\ $\si{\milli\kelvin}. The magnetic field $B$ was applied out of plane and perpendicular to the junction. Four terminals at the chip level measure the JJ in a two-terminal geometry (which removes the series resistance of the wires down the cryostat). The differential resistance $R=\frac{dV}{dI}$ is measured with lock-in amplifier at a frequency of 19 \si{\hertz} and excitation of $1\, \si{\nano\ampere}$ as a function of back gate voltage ($V_{BG}$) and the applied DC current bias ($I_{dc}$). Fig.~\ref{fig1}(b) shows the $V_{BG}$ dependence of the critical current. The charge neutrality point (CNP) is found to be at $V_{BG}=-0.75$ \si{\volt}, corresponding to the lowest critical current and the largest normal state resistance. Fig.~\ref{fig1}(c) shows the Fraunhofer pattern $R(B)$ at the CNP. The regularity of the pattern implies the supercurrent is distributed uniformly along the contacts at the CNP.

\begin{figure}
    \label{fig1}
    \includegraphics[width=2.75in]{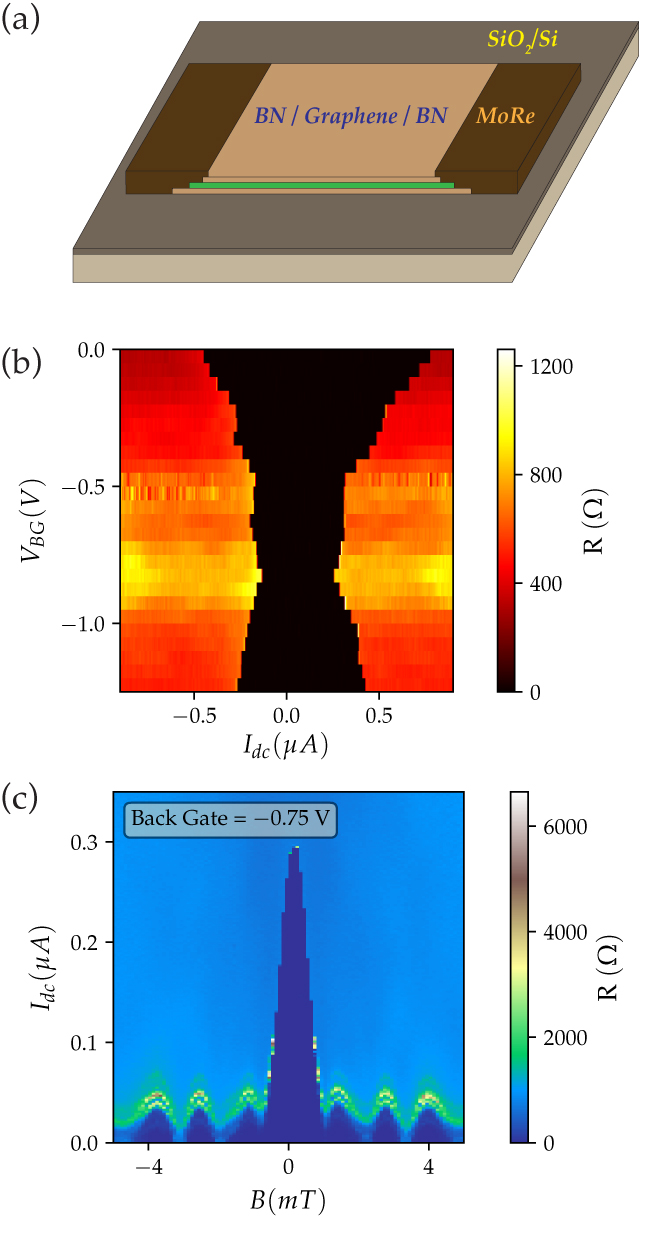}
    \caption{(a) A schematic of the h-BN encapsulated graphene device. Side contacts to graphene are made with molybdenum-rhenium (MoRe) as the superconductor. The device is fabricated on a SiO$_2$/Si substrate and can be backgated. (b) Dependence of the critical current on $V_{BG}$ near the CNP, which occurs at $V_{BG}=-0.75$ \si{\volt}. (c) Magnetic diffraction (Fraunhofer) pattern for the junction at CNP. The periodicity implies a junction area of about 1.5 \si{\micro \meter^2}.}
      \label{fig1}
\end{figure}

\section{Measurement Results}
In order to probe the phase dynamics of the JJ, $R$ is measured as a function of $I_{dc}$ and RF power. When junctions are periodically driven, phase locked solutions produce a a fixed voltage across the JJ. These are known as Shapiro steps \cite{shapiro_effect_1964}, and show up as zero differential resistance in a lock-in measurement.  The measurement of the Shapiro diagram yields information about the CPR and has been used previously to study weak links of topological materials~\cite{rokhinson_fractional_2012,wiedenmann_4_2016, veldhorst_josephson_2012, pribiag_edge-mode_2015, Snyder2018, Wang2018, Yu2018, Calvez19}. 

\begin{figure*}[t!]
    \centering
    \includegraphics[width=7in]{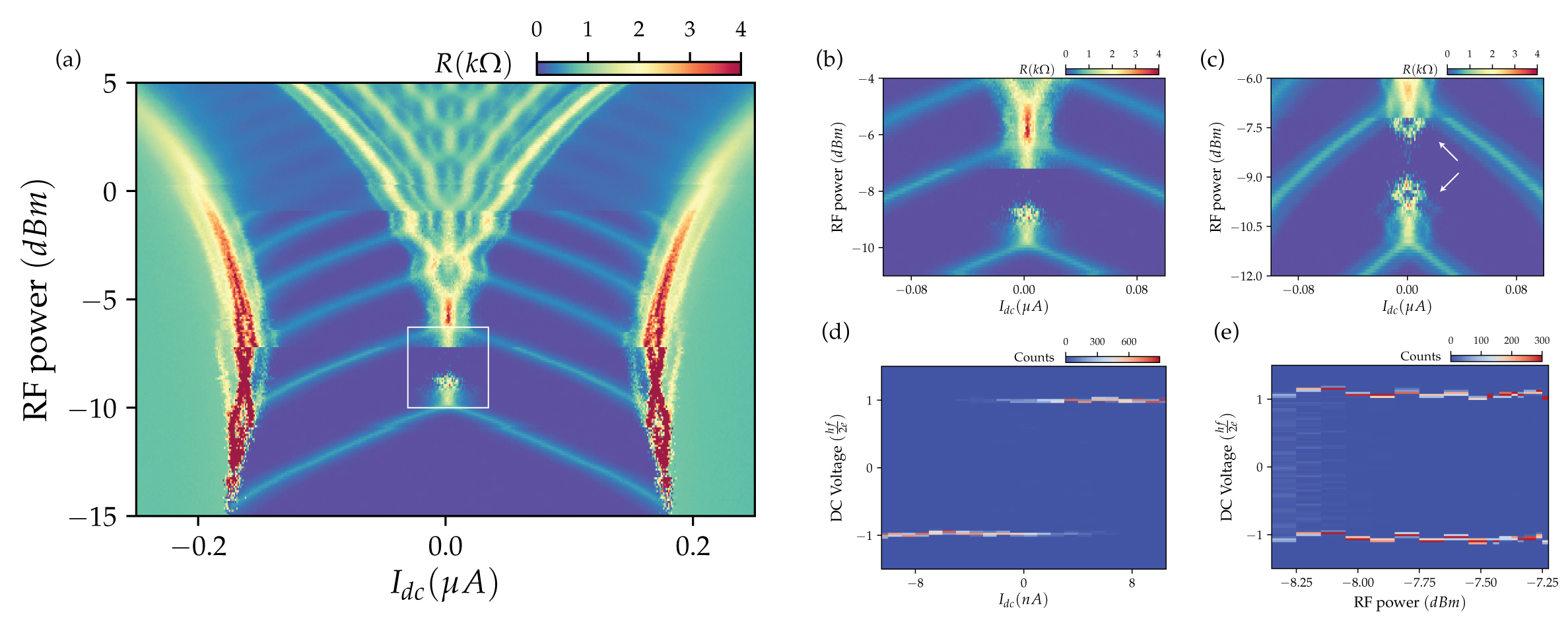}
    \caption{(a) Shapiro diagram at $f = 6.350$ \si{\giga\hertz}, $V_{BG}$ = $-0.75$ \si{\volt} and $B$=0. The first elongated note at the lowest RF power is also broken (white box). (b) A close-up of the broken node region of (a). (c) Shapiro diagram at $f = 6.350$ \si{\giga\hertz}, $V_{BG}$ = $-0.75$ \si{\volt} and $B$=0.3\,mT. The broken node shifts in position and extra featues appear are observed (white arrows). (d, e) Direct DC voltage, binned in regular time intervals, around the broken node as a function of $I_{dc}$ and RF power. The measured voltages values cluster around two voltages and overlap, indicating a bistability between the $\pm$1 Shapiro steps. }
    \label{fig2}
\end{figure*}

\subsection{Shapiro Diagram and Broken Node}
Fig.~\ref{fig2}(a) shows a Shapiro diagram at RF frequency ($f$) of $6.350$ \si{\giga\hertz} and $V_{BG}=-0.75$ \si{\volt}. We observe regions of zero differential resistance corresponding to Shapiro steps separated by resistive transitions between steps. This Shapiro diagram is distinct in two principal ways from previous measurements on graphene \cite{heersche_bipolar_2007} and other materials \cite{rokhinson_fractional_2012,wiedenmann_4_2016, veldhorst_josephson_2012, pribiag_edge-mode_2015, Snyder2018, Wang2018, Yu2018, Calvez19}. The first is the resistive transitions are elongated node-like regions, not single points that are expected for the Bessel function step width as a function of RF power~\cite{PaternoBarone}. These elongated nodes are a generic feature of underdamped JJs (explained below) and can be used to infer underdamped behavior even when samples lack hysteresis in measurements of a $I-V$  curve. Second, the first node separating the $\pm 1$ steps at $I_{DC}$=0 disappears (indicated by the white box) for certain RF powers. Switching behavior on a timescale slower than the integration time for a lock-in measurement can lead to such a feature. Since the integration time used was around a second, this implies a switching timescale on the order of seconds. Fig.~\ref{fig2}(b) shows a closeup of the broken node region. In addition to the broken node, the details of resistive transition is dependent on the junction parameters. Fig.~\ref{fig2}(c) shows the changes when $I_{DC}$ is reduced by $B$, where additional structure in $R$ is observed (white arrows). For a more extensive visualization of the changes to the broken node region in response to small changes in $f$, $V_{BG}$ and $B$, see the Appendix A.

To investigate the broken node region further, we switch to direct DC voltage measurements as a function of time. Figs.~\ref{fig2}(d,e) show a histogram of the voltage values measured as function of $I_{dc}$ and RF power. For a given $I_{dc}$ or RF power, we measure the DC voltage $V(t)$ at regular time intervals. These values are then binned. The counts are reported as the color in Figs.~\ref{fig2}(d,e). When the count distribution is bi-modal, a bistability between the two values can be inferred. We clearly observe a bistability between the two $\pm$1 Shapiro steps at the broken node as a function of $I_{dc}$ [Fig.~\ref{fig2}(d)] and RF power [Fig.~\ref{fig2}(e)].

\subsection{Bistability and Switching Time}
\begin{figure}[h!]
    \centering
    \includegraphics[width=\linewidth]{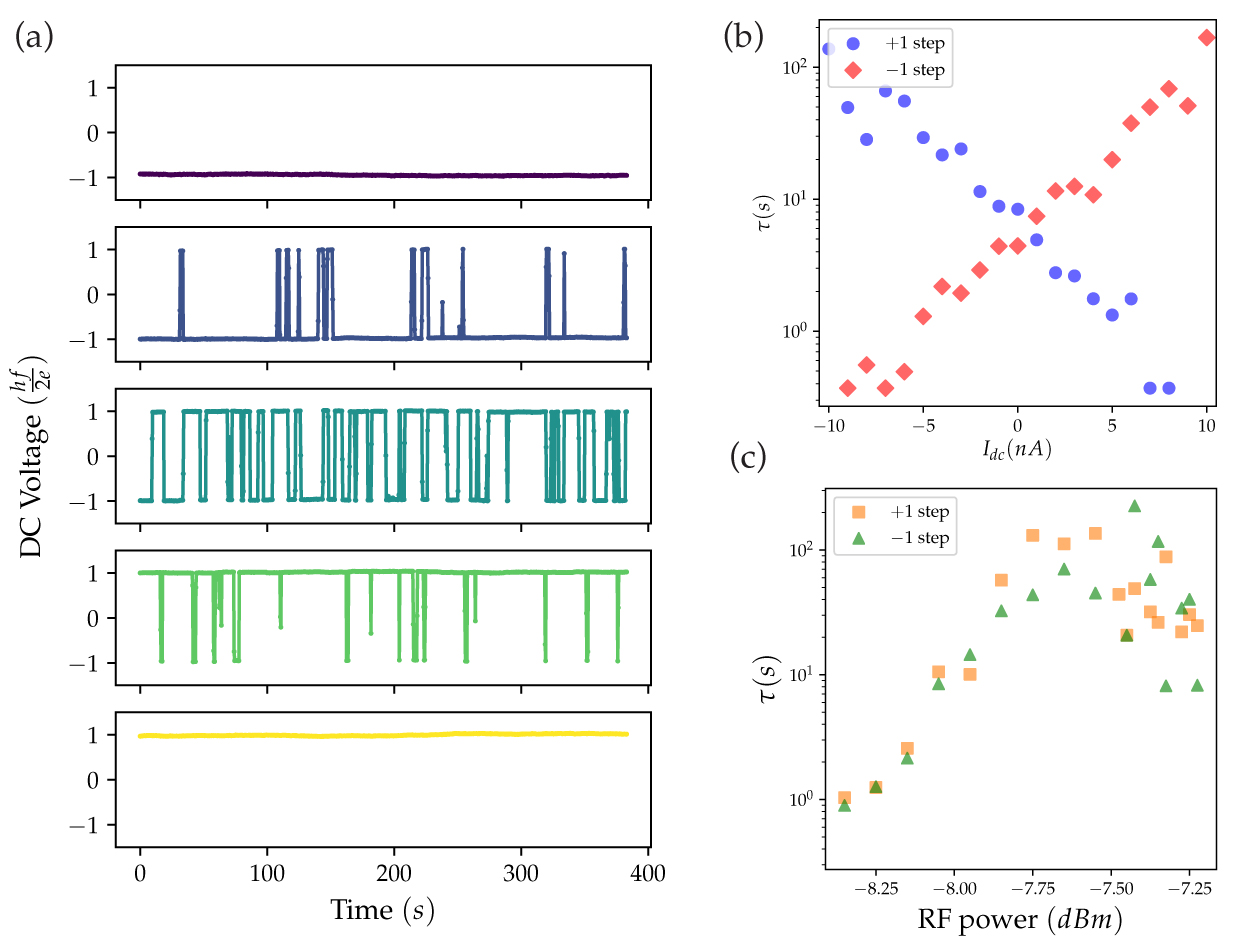}
    \caption{(a) DC Voltage measured as a function of time, measured at $I_{dc}$ values of $\{-14,-3,1,5,14\}$ \si{\nano\ampere} (top to bottom). A change from being entirely on $-1$ to $+1$ Shapiro step is seen. Near zero current bias, the voltage stochastically switches between the two steps. A long switching timescale, of order 10 seconds, is observed. (b, c) Dependence of the switching timescale $\tau$ on $I_{dc}$ and RF power measured in the region of the broken node.}
     \label{fig3}
\end{figure}

Fig.~\ref{fig3}(a) shows the evolution of  $V(t)$ with $I_{dc}$ from negative to positive values. For large negative values ($\sim -10 \si{\nano\ampere}$) of $I_{dc}$, the junction is phase-locked to the $-1$; the same occurs for large positive values of $I_{dc}$. However, in the transition region between $-1$ and $+1$, we see a bistable region. $V(t)$ randomly switches between $\pm 1$ steps. $\tau$ is defined as the average time $V(t)$ spends on a particular step. $\tau$ is extraordinarily long -- on the order of seconds -- much longer than any junction timescales. For comparison, a typical timescale associated with JJ dynamics is the characteristic frequency of the junction, defined as $f_c=\frac{2 e I_c R}{\hbar}$, where $I_c$ is the critical current and $R$ is the normal state resistance.  $f_c$ for this junction is $\sim$ 50 \si{\giga\hertz}. Fig.~\ref{fig3}(b) shows the dependence of $\tau$ on $I_{dc}$.  $\tau$ changes by almost three orders of magnitude in the bistable region. A similar scaling of $\tau$ now as a function of RF power is shown in Fig.~\ref{fig3}(c). Note the log scale on the $\tau$ axis in Figs.~\ref{fig3}(b,c).

Such scaling behavior is unexpected. One would expect that the dynamics smoothly change between $\pm 1$ steps. The scaling with RF power also shows a maximum in $\tau$ at around $-7.4$ dBm in Fig.~\ref{fig3}(c) and quickly decays away from this maximum.  These observations point out that it is crucial to understand exact nature of the phase locked solutions though a model, which we do in the following section.

\section{RCSJ Model}
A model for the evolution of $\phi(t)$ under external DC and RF drives can be used to understand the Shapiro diagram. A simple circuit model for JJs, called the resistively and capacitively shunted junction (RCSJ) model [Fig.~\ref{fig4}(a)]~\cite{stewart_currentvoltage_1968,mccumber_effect_1968}, has been widely used for understanding phase dynamics in JJs. In this section, we describe the parameters for our model and the modifications required when graphene serves as the weak-link. $\phi$ can be related to experiment by calculating voltage across the junction using the Josephson relation $V = \frac{\hbar }{2e} \frac{d\phi}{dt}$ \cite{josephson_possible_1962}. 

\subsection{Model Definition}
\begin{figure}[H]
   \centering
    \includegraphics[width=\linewidth]{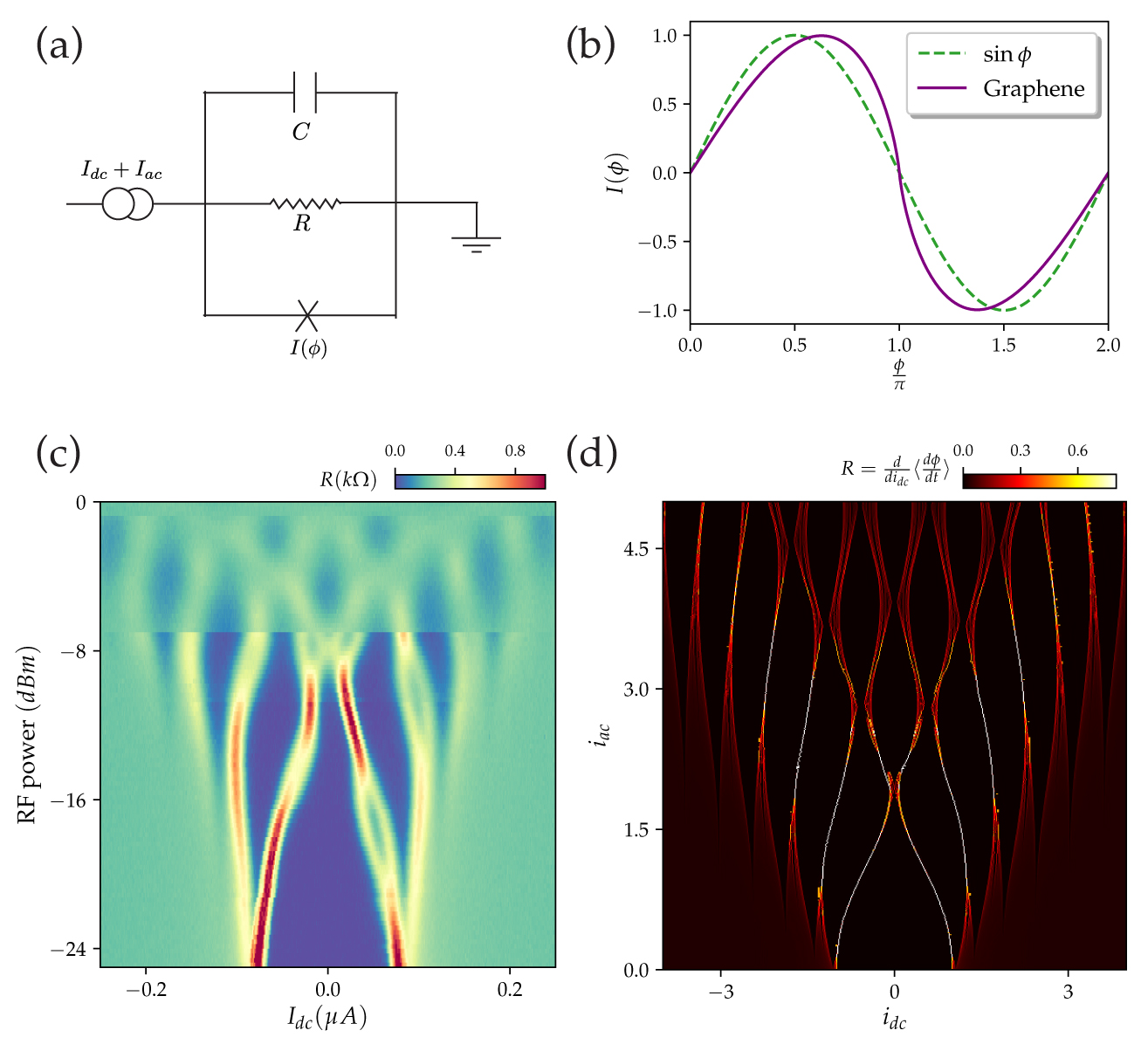}
     \caption{(a) Circuit diagram of the RCSJ model. A supercurrent whose phase dependence is given by CPR is assumed to flow in parallel to a resistor and a capacitor. External DC and AC currents can be used to drive the junction. (b) A comparison between graphene and a sinusoidal CPR. (c) Shapiro diagram measured at $6.350$ \si{\giga\hertz}, $V_{BG}$ = $+0.25$ \si{\volt} and magnetic field set to 0.7 \si{\milli\tesla}. (d) Simulated Shapiro diagram for $Q = 1$ and $\Omega = 0.6$ which qualitatively matches the features seen in (c).   }
    \label{fig4}
\end{figure}

In junctions with a conducting weak link, the super-current flow is mediated by Andreev bound states (ABS), formed by Andreev reflection at the two weak link-superconductor interfaces \cite{beenakker_colloquium:_2008,pillet_andreev_2010}. The position of ABSs depends on $\phi$, the electron/hole dispersion in the weak link region and on the coupling strength between the superconductors and the weak link. For a short and wide ballistic graphene junction at the Dirac point, the CPR is \cite{titov_josephson_2006},

\begin{align}
I(\phi) &= \frac{e \Delta}{\hbar} \frac{2 W}{\pi L} \left( \cos{\frac{\phi}{2}} \right) \tanh^{-1}\left(\sin{\frac{\phi}{2}}\right)   \\
&= I_c \frac{2}{1.33} \left( \cos{\frac{\phi}{2}} \right) \tanh^{-1}\left(\sin{\frac{\phi}{2}}\right) 
\end{align}

\noindent where $\Delta$ is the superconducting order parameter, $W$ is junction width perpendicular to current flow and $L$ is the junction length. The numerical factor of $\frac{2}{1.33}$ ensures that the maximum as a function of $\phi$ is $I_c$. See Fig.~\ref{fig4}(b) for a plot of this CPR.

Once the CPR is established, the phase dynamics can be modeled by RCSJ model and the governing equation is:

\begin{align}
    \frac{\hbar C}{2 e} \frac{d^2\phi}{dt^2} + \frac{\hbar}{2 e R} \frac{d\phi}{dt} + I(\phi) = I_{dc} + I_{ac}\sin(\omega t)
\end{align}
where $C$ is the junction capacitance, $R$ is the normal state resistance, and $I_{dc}$ and $I_{ac}$ are the externally applied DC and AC currents. The equation can be cast into dimensionless form by transforming $t \rightarrow t \sqrt{\frac{2 e I_c}{\hbar C}}$:

\begin{align} \label{eqn-rcsj-model}
    \frac{d^2\phi}{dt^2} + \frac{1}{Q} \frac{d\phi}{dt} + i(\phi) = i_{dc} + i_{ac} \sin(\Omega t) 
\end{align}
where $Q = \sqrt{\frac{2 e I_c R^2 C}{\hbar}}$ is called the quality factor and $\Omega = \omega \sqrt{\frac{\hbar C}{2 e I_c}}$ is the normalized frequency. Depending on the value of the capacitance and consequently the Q value, the junction can be overdamped ($C \rightarrow 0 \therefore Q \rightarrow 0)$ or underdamped ($Q \gtrsim 1$). 

An equivalent first order autonomous form $\frac{d y}{dt} = F(y)$ can be obtained by defining $y = (\phi,\frac{d\phi}{dt},\Omega t)$. This implies that the phase space for this system has to be three-dimensional, and hence solutions are not limited to fixed points and limit cycles, but strange attractors and chaotic effects are possible (Poincare-Bendixson theorem \cite{guckenheimer_nonlinear_2002,ott2002chaos}). On the other hand, in the limit $C \rightarrow 0$, the second order term vanishes and we are left with two-dimensional phase space, where the dynamics is restricted to fixed points and limit cycles. In the $C \rightarrow 0$ limit, we recover the overdamped or resistively shunted junction (RSJ) model, whose dynamics are well-understood. See the Appendix B for a discussion on Shapiro diagrams in the overdamped limit.

The flow on $y$ defined by $F(y)$ has a Jacobian equal to $\frac{-1}{Q}$ implying a dissipative system. It can be shown that in the presence of dissipation, the dynamics of a JJ with a sinusoidal CPR is equivalent to the driven pendulum. This correspondence to nonlinear dynamics of driven-damped pendulum has been explored before in simulations and connections to JJ experiments have been drawn, as we elaborate on in the next subsection. 

\subsection{Previous Work on Underdamped, Driven Chaotic Systems}

Early motivation to consider the nonlinear dynamics of Josephson junctions arose from the large values of electric noise observed in experiment. It was observed that JJ-based parametric amplifiers resulted in broad-band voltage fluctuations~\cite{Taur1977} with a noise temperature on the order of $10^4$ K, which was ascribe to the chaotic behavior of Eq. \ref{eqn-rcsj-model} \cite{huberman_noise_1980}. It was suggested that this arises from the presence of a strange attractor in phase space for certain set of parameters characterizing the junctions. 

Experimentally, Poincare maps and bifurcations present in the parameter space were first measured with phase-locked electrical loops whose dynamics are similar to JJs and the driven simple pendulum equation \cite{cirillo_bifurcations_1982}. A similar system was used to investigate symmetry breaking in solutions of the equation as a period-doubling route to chaos. Chaotic and intermittency was reported in measurements of bifurcation cascades of phase-locked loops \cite{dhumieres_chaotic_1982}. Intrinsic and noise-induced intermittency due to crisis has been been studied with numerical simulations. The authors reported an intermittent switching behavior between positive and negative voltage solutions with characteristic power-law type relation for the switching timescale \cite{gwinn_intermittent_1985}.   The basin boundary for these solutions was found to be fractal. 
 
Extensive numerical simulations  were carried to investigate the chaotic behavior of JJs and the effect of noise \cite{kautz_ac_1981,kautz_chaotic_1981}. In experiments with a Josephson junction, a correlation was found between the noise and the fractal dimension of the boundary between periodic solutions \cite{gwinn_fractal_1986,iansiti_noise_1985}.  Despite the extensive numerically simulations of the driven pendulum system, experimental verification remain scarce and are restricted to qualitative comparison \cite{octavio_chaos_1984,iansiti_noise_1985,hu_noise_nodate, Davidson1986}. 
\begin{figure*}[t]
     \centering
     \includegraphics[width=7in]{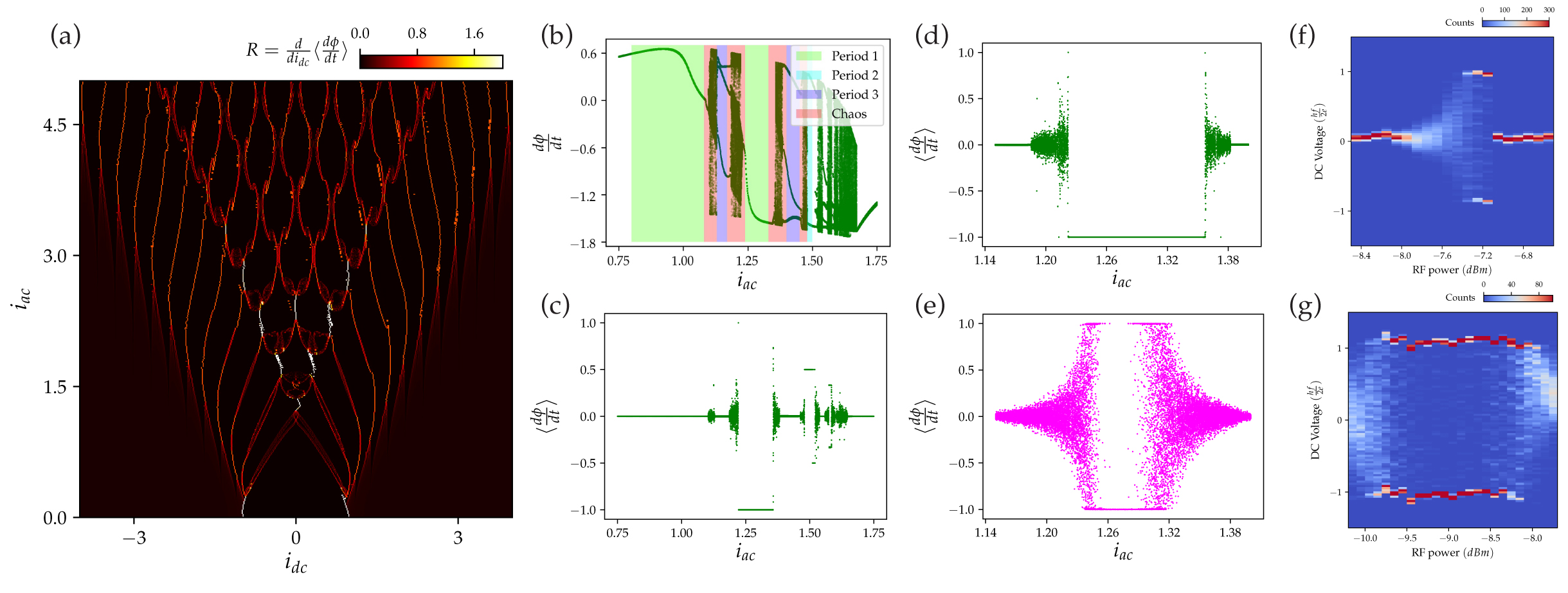}
     \caption{(a) Shapiro diagram simulated for $Q = 1.5$ and $\Omega = 0.5$, and initial condition $(\phi(t=0),\dot{\phi}(t = 0)) = (0,0)$. Contours separating two steps have an intricate structure as opposed to point touchings. In particular, along the $i_{dc} = 0$ line, the contour separating $\pm 1$ steps in elongated. (b) Poincare map $\phi(nT)$ for same parameters as a with $i_{dc} = 0$ and $i_{ac}$ varied in a range around the elongated node separating $\pm 1$ steps. Periodic attractors with different periods (corresponding to number of points in the Poincare map for a given $i_{ac}$), as well as chaotic windows separating them are observed. (c) DC limit of the Poincare map with a window size of 200 periods. (d) A fine scan of the period 1 attractor of the $+1$ step. (e) Effect of noise on the attractor from (d). Both $\pm 1$ steps are possible solutions and a bistable region emerges. (f, g) Experimentally measured histograms of $V(t)$ values as a function of RF power at the broken node for ($f$, $B$)=($6.322\, \si{\giga\hertz}, 0\, \si{\milli\tesla})$ (f) and ($6.350\, \si{\giga\hertz},0.5\, \si{\milli\tesla})$ (g).}
     \label{fig5}
\end{figure*}

\subsection{Simulation Results for the Underdamped RCSJ Model}
We now describe the simulation of a Shapiro diagram under the RCSJ model. The voltage across the junction is given by $V(t) = \frac{\hbar}{2e} \frac{d\phi}{dt}$. Since the voltage is measured in the DC limit, the average voltage is given as:

\begin{align}
    \frac{2 e}{\hbar} \langle V \rangle_n &= \langle \frac{d\phi}{dt} \rangle\Big|_n \\
                      &= \frac{1}{nT} \int_0^{nT} \frac{d\phi}{dt} dt \\
                      &= \frac{\phi(nT) - \phi(0)}{nT}
\end{align}
where $T = \frac{2\pi}{\Omega}$ is the drive period and n is large number to average out transient behavior. For a phase locked solution, $\phi$ advances by an integer number of $2\pi$ period, and $\langle \frac{d\phi}{dt} \rangle$ at the $m^{th}$ Shapiro step is given by $m \Omega$. The DC limit is achieved when $n \rightarrow \infty$. It is also useful to define a window average for a a fixed number of periods which we denote as $\langle V \rangle_n = \langle \frac{d\phi}{dt} \rangle\Big|_n$. We will use both the DC limit and fixed window averages in the following sections. The values of  $\langle V \rangle_n = \langle \frac{d\phi}{dt} \rangle\Big|_n$ are normalized in units of $\Omega$ so that $m^{th}$ Shapiro step corresponds to $\langle V \rangle = \langle \frac{d\phi}{dt} \rangle\Big|_n = m$. It is also important to consider the intial conditions for solving the ODE given by Eq. \ref{eqn-rcsj-model}. Since, it is of second order, a unique solution requires two initial numbers and we use the initial condition $(\phi(t=0),\dot{\phi}(t = 0)) = (0,0)$ unless otherwise stated.
In order to understand the effects of noise, a Wiener process in time is added to the RHS of Eq. \ref{eqn-rcsj-model}. This models a Gaussian white current noise source with zero mean. As a result, the RCSJ model with noise becomes a stochastic differential equation (SDE) as follows: 

\begin{align} \label{eqn-rcsj-noise}
    \frac{d^2\phi}{dt^2} + \frac{1}{Q} \frac{d\phi}{dt} + i(\phi) = i_{dc} + i_{ac} \sin(\Omega t) + \sigma_n \eta(t)
\end{align}
where $\eta(t) dt$ is a Weiner process and $\sigma_n$ controls the noise strength. It is useful to think of $\sigma_n^2$ as being proportional to a energy or temperature scale, which would hold exactly if the noise sources were due to Johnson-Nyquist noise in resistors. In simulations with noise, we integrate this SDE with the Euler-Maruyama method. 

To understand our experimental results, we consider junctions in the weakly underdamped limit where $Q \approx 1$. Since the junction capacitance is not measured directly, we indrectly infer that the value of $Q$ must be close to 1. Fig.~\ref{fig4}(c) shows a Shapiro diagram measured at $6.350$ \si{\giga\hertz}, $V_{BG}$ $ = +0.25$ \si{\volt} and magnetic field of $0.7$ \si{\milli\tesla}. We see that results are different from those at CNP and the broken node is not present. Fig.~\ref{fig4}(d) shows a simulated Shapiro diagram with $Q = 1$ and $\Omega = 0.6$; qualitative reproduction of the measured features is observed. In particular, the contours separating the steps, meet in points ($\pm 1$ steps) or show a small gap between them. Since the values for $V_{BG}$ and $B$ chosen in Fig.~\ref{fig4}(c) cause a reduction in $R$ and $I_c$ and hence a reduction in $Q$, we expect the dynamics at CNP and zero magnetic field to be described by a model with $Q > 1$.

Fig.~\ref{fig5}(a) shows a simulated Shapiro diagram for $Q = 1.5$ and $\Omega = 0.5$. As opposed to diagrams in the overdamped limit (Appendix B), there are many distinctive features. The resistive contours for the steps meet in elongated notes as opposed to single points, similar to the experimentally observed features [Fig.~\ref{fig2}(a)]. For a more detailed comparison between the fine features of the measurements and simulations, see the Appendix C. 

\subsection{Discrete symmetry of underdamped RCSJ model}
We now focus on the explanation on the broken node feature. Since the node occurs on the $I_{dc} = 0$ line, we consider the following equation:

\begin{equation}\label{eqn:rcsj-idc-zero}
    \frac{d^2\phi}{dt^2} + \frac{1}{Q} \frac{d\phi}{dt} + i(\phi) = i_{ac} \sin(\Omega t). 
\end{equation}

\noindent Eqn.~\ref{eqn:rcsj-idc-zero} has a discrete symmetry among the possible solutions. For $\phi \rightarrow -\phi$ and $t \rightarrow t + \frac{\pi}{\Omega}$, the equation is left invariant. Note that, we also need $i(\phi) = -i(-\phi)$ and this holds from time-reversal symmetry \cite{golubov_current-phase_2004}. This symmetry implies,

\begin{equation}
    \phi(t) \text{ is a solution} \implies -\phi(t +\frac{\pi}{\Omega}) \text{ is a solution}.
\end{equation}
Under this operation, for given values of $Q, \Omega$ and $i_{ac}$, if $\langle V \rangle = m$ is possible, so is $\langle V \rangle = -m$. Hence if $m \neq 0$, it is possible for the system to have two distinct solutions with an \emph{inequivalent observable} (the DC voltage) in a measurement. Which of these two phase locked solution is achieved depends on the initial conditions  $(\phi(t=0),\dot{\phi}(t = 0))$. It is convenient to think of a basin structure \cite{ott2002chaos} for each of the two solutions, where each initial condition eventually settles into one of these two possible solutions. We wlll later see how an interplay between the basin structure and the noise level can lead to the switching characteristics measured in Fig.~\ref{fig3}(b,c).

\subsection{Poincare Maps and Bistability}
A handy construction for elucidating the behavior of a dynamical system is a Poincare map or section. For a dynamical system, it is the intersection of a periodic orbit with a lower-dimensional subspace, transverse to the flow of the system. In our case, for the RCSJ model, we consider $(\phi(t),\frac{d\phi}{dt}(t))$ for $t =n\frac{2\pi}{\Omega}$ where $n$ is an integer.  

In Fig~\ref{fig5}(b), the Poincare map for $\frac{d\phi}{dt}$ with parameters $Q = 1.5$ and $\Omega = 0.5$ is plotted. $i_{dc}$ fixed to 0 and $i_{ac}$ is varied in a range around where the $\pm 1$ steps meet. The Poincare map has a rich structure with periodic attractors, separated by windows where the dynamics causes $\frac{d\phi}{dt}$ to spread uniformly over a range of values. The behavior in these windows is similar to the periodic doubling bifurcation cascade in the logisitc map \cite{ott2002chaos} and we refer to these windows as chaotic windows. In general, the phase space structure of the model along $i_{ac}$ axis consists of periodic attractors separated by chaotic windows. 

The phase dynamics $\phi(t)$ is challenging to measure directly in an experiment and the only quantity accessible is the DC voltage. To this end, we now discuss Poincare map with an average over a fixed window. We simulate dynamics for 10000 periods, average over 200 consecutive periods and plot the values taken by the window averaged $\langle \frac{d\phi}{dt} \rangle\Big|_{200}$ for each value of $i_{ac}$. We call this the DC limit of the Poincare map and it is pragmatic in explaining our observations. Fig.~\ref{fig5}(c) shows the DC limit of the map from Fig.~\ref{fig5}(b). The chaotic windows average out to noisy regions while the periodic attractors attain their DC values. An attractor to the $+1$ step is found to occur for $i_{ac} \in (1.23,1.33)$. Fig.~\ref{fig5}(c) shows a closeup of the DC Poincare map of this attractor. As $i_{ac}$ is increased in this range, we move from being on an attractor with $\langle \frac{d\phi}{dt} \rangle = 0$ to $\langle \frac{d\phi}{dt} \rangle = +1$ and back to 0 again, separated by noisy regions due to the chaotic windows. 

The symmetry argument made in the last section, implies that the attractor in the range $i_{ac} \in (1.23,1.33)$ also has solutions corresponding to the $-1$ step. Which attractor the system settles in depends on the initial conditions. However, the presence of noise removes any bias corresponding to the initial conditions and causes the system to switch between $\pm 1$ attractors. Fig.~\ref{fig5}(e) show this effect, where a noise term with $\sigma_n = 4 \times 10^{-2}$ is included. It is useful to recall the basin structure, where each set of initial conditions lead to one of the attractors. The presence of noise causes a drift in initial conditions after every period and so it becomes possible for the solution to switch from one attractor to another. 

Fig.~\ref{fig5}(f,g) show the histogram of experimentally measured $V(t)$ values measured as a function of RF power in the broken node region $(f,B)$ = $6.322$ \si{\giga\hertz}, $B$=0 (f) and $6.350$ \si{\giga\hertz}, $B$=0.5 \si{\milli\tesla} (g). We see that the observed behavior of the DV voltage in the broken node region is qualitatively similar to the dynamics in the DC limit of Poincare maps in Fig.~\ref{fig5}(b-e). An attractor to $\pm 1$ steps is sandwiched between attractors to the $0$ step separated by noisy regions corresponding to the chaotic windows leads to the broken node feature observed in the experiment. 

\begin{figure*}
    \centering
    \includegraphics[width=6in]{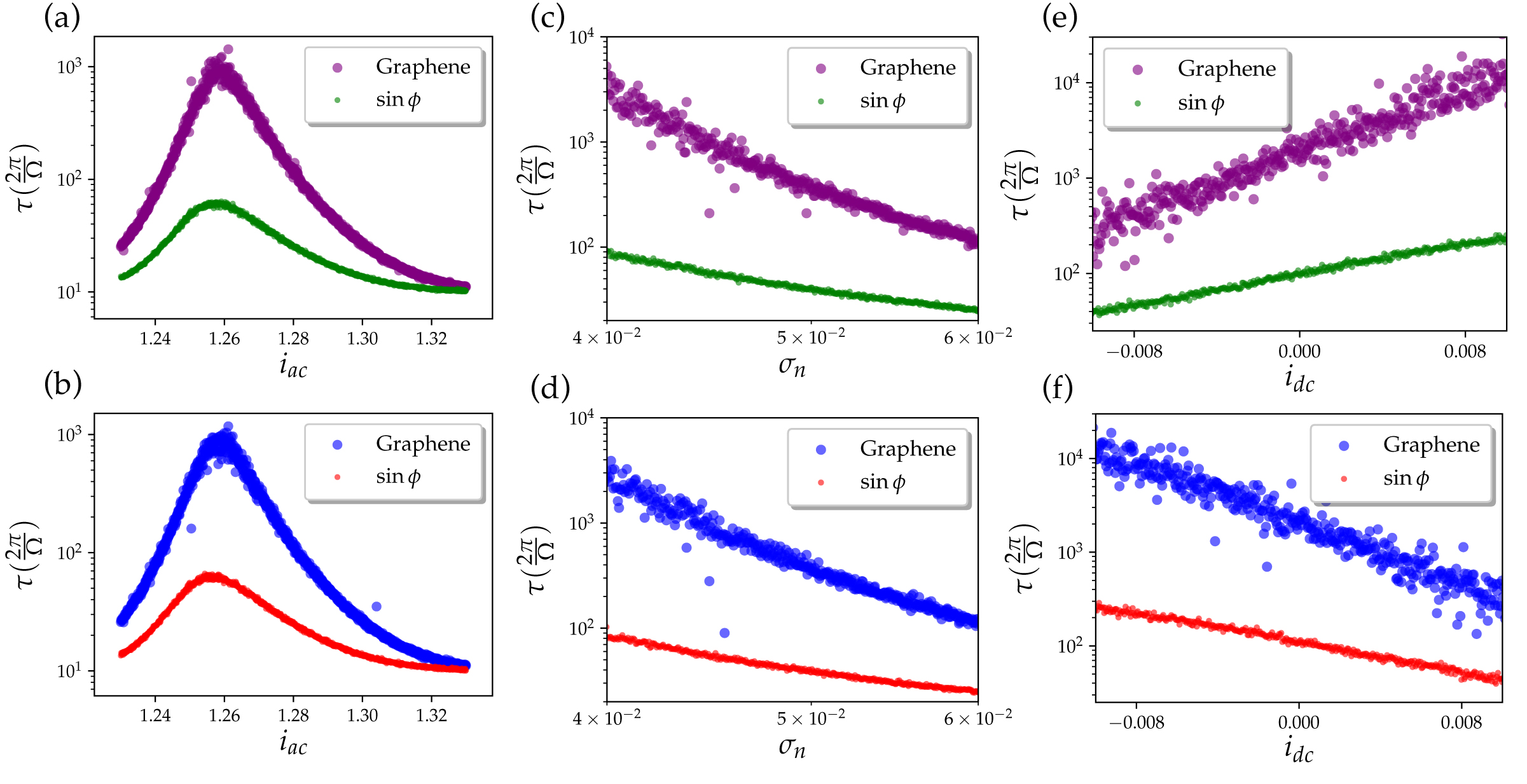}
    \caption{(a,b) The simulated evolution of switching time on a bistable attractor of the $+1$ (a) and $-1$ (b) steps for $i_{dc} = 0$ and $\sigma_n = 4 \times 10^{-2}$. (c,d) $\tau$ as a function of $\sigma_n$ at $i_{dc} = 0$ and $i_{ac} = 1.28$ for the $+1$ (c) and $-1$ (d) steps. For low noise levels, $\tau$ becomes very large and decreases with increased noise. (e,f) Effect of $i_{dc}$ on $\tau$ for the $+1$ (e) and $-1$ (f) steps. As $i_{dc}$ is increased from negative to positive values, the solution prefers to stay more on $+1$ step than on $-1$ step; the corresponding changes in $\tau$ are evident. In all the plots, the effect of the graphene to sinusodinal CPR is compared. Though both CPRs show a scaling of $\tau$ with $i_{ac}$, the overall magnitude is larger for the graphene CPR. A window size of 10 periods was used to take the DC limit and the system was simulated for a total of $10^5$ periods.}
    \label{fig6}
\end{figure*}

\subsection{Scaling of switching time}
Having understood the origin of bistability in simulation, we now focus on the changes in switching time $\tau$ as a function of DC and AC currents. We reported that $\tau$ undergoes changes over orders of magnitude when $I_{dc}$ and RF power are changed in Fig.~\ref{fig3}(b,c). We now reproduce similar scaling in simulation. Fig.~\ref{fig6} shows the changes in $\tau$ as a function of $i_{ac}$, noise level $\sigma_n$ and $i_{dc}$. We use the same set of parameters and the $\pm 1$ attractor discussed in the last section.

In Fig.~\ref{fig6}(a,b), we plot the switching time $\tau$ for the $+1$ and $-1$ steps respectively. We also compare the effect of the graphene CPR to a simple $\sin(\phi)$ CPR. While fixing all other parameters, the CPR of graphene causes the switching time to be much larger. This feature also occurs with the scaling of switching time as a function of noise level and $i_{dc}$ [Fig.~\ref{fig6}(b-e)]. This can be attributed to the graphene CPR leading to an effective potential for $\phi$ which has steeper and deeper minima, making escape for a given noise level less likely. This increase in $\tau$ resulting from the CPR of graphene serves as an explanation why the bistability might have been harder to directly measure in other underdamped systems.

The presence of chaotic windows for a range of parameters complicates an analytic approach that might elucidate the origin of the observe scaling. Lacking any exact theoretical basis, we now give a heuristic argument for the scaling of the switching time. It is helpful to recall the basin structure picture, where the noise causes a drift of initial conditions, allowing for the solution to switch from the basin of one attractor to another. Formalizing this notion, suppose, the noise causes a diffusion at time $t$  as $\Delta\phi(t) \sim (\sigma_n \sqrt{t})^\nu$, where the term $\sigma_n \sqrt{t}$ has been chosen because it is dimensionless and $\nu$ is a diffusion exponent. $\nu = 1$ corresponds to standard Brownian motion, where net displacement is proportional to $\sqrt{t}$. Suppose $f(i_{ac},i_{dc})$ represents a scale in the basin structure; when initial conditions are changed by this scale, a switch from one attractor to another occurs. We can therefore write a condition for the switching timescale $\tau$:

\begin{equation}
    \Delta \phi(\tau) \sim f(i_{ac},i_{dc}).
\end{equation}
Using the diffusion expression for $\Delta \phi$, the following expression is obtained:

\begin{equation}\label{eqn:scale-model}
    \tau \sim \frac{r(i_{ac},i_{dc})^{1/\nu}}{\sigma_n^2}.
\end{equation}
The details of the function $r$ depends on the exact nature of the CPR, noise level $\sigma_n$ as well as the parameters $Q$ and $\Omega$. Though we don't have a way to calculate $ f(i_{ac},i_{dc}$, Eq.~\ref{eqn:scale-model} show why such relationships might exist in the first place. Further theoretical work is necessary to elucidate the scaling relationship of $\tau$ in the model.

\section{Conclusion and Outlook}
In summary, we have reported on the AC Josepshon in h-BN encapsulated graphene JJs. A number of features distinct from previous measurements from un-encapsulated graphene junctions as well as other systems have been observed. The resistive contours separating two steps meet in elongated nodes as opposed to points expected from RSJ model. Incorporating the capacitance and considering the full RCSJ model in the weakly underdamped limit leads to a provide an origin of these elongated nodes. Moreover, we observe a bistability at the first elongated node. The switching timescale is extraordinarily long, on the order of seconds, and is much slower than any timescale relating to the junction dynamics. A change in $\tau$ over almost three orders of magnitude is observed with changes in external applied DC currents and RF power.

Simulation of the RCSJ model reveals a rich dynamical structure where periodic attractors are interleaved with chaotic windows. In the DC limit, this leads to attractors separated by noisy regions. In particular, we show the presence of an attractor for $\pm 1$ steps whose dynamics in the presence of noise is similar to the bistability observed in the broken node region. Simulation the switching time in the bistable region as a function of $i_{ac}$, $i_{dc}$ and noise level offers an explanation of the scaling behavior observed in the experiment. We also show how the graphene CPR effects the overall magnitude of the switching timescale. We offer a heuristic argument for the existence of such scaling from the interplay between the basin structure and the noise level.

A driving theme in contemporary condensed matter research is the realization of Majorana particles and topologically non-trivial ground states \cite{wilczek_majorana_2009,wilczek_majorana_2014,lutchyn_majorana_2010,oreg_helical_2010}. One of the principal ways to detect topology in these systems is the $4\pi$ Josephson effect and has been used to this effect in certain experiments \cite{rokhinson_fractional_2012, wiedenmann_4_2016, laroche_observation_2019}. Yet, chaotic behavior can mimic this effect: simulation above show that period doubling is possible for certain values of $i_{ac}$. Our work points to a key feature which can indicate the potential for chaotic behavior (i.e. underdamped) in a JJ -- the elongation of nodes in a Shapiro diagram. This seems to be more accurate that the standard means of inferring underdamped junctions, hysteresis, which may be absent in slightly underdamped ($Q \sim1$) JJs. For example, work on HgTe JJs estimate an overdamped junction from device parameters~\cite{Oostinga2013} and as a result rule out the possibility chaotic behavior in the JJs~\cite{wiedenmann_4_2016}. Yet observe elongated nodes seem to contradict this assessment.  Further, gaps in resistive nodes have been ascribed to a 4$\pi$-periodic contribution to the CPR~\cite{Wang2018, Calvez19}. In this work, we observe this gap to arises entirely from nonlinear behavior in the junction [Figs. 4(c,d)].  As the coupling between the superconductor and the material forming the weak link is improve and as the mobility of the weak link is increase, larger values of $I_C$ and $Q$ are expected. Hence, a systematic delineation of phase dynamics of driven JJs is crucial for using the AC Josephson effect as a probe of topology and exotic physics.

The underdamped RCSJ model is dissipative with a strength parametrized by $\frac{1}{Q}$. In the limit $Q \gg 1$, the RCSJ model becomes a conservative Hamiltonian system. We have let $\phi$ be a classical variable, though it is possible to quantize the system. It would be interesting to study the quantum dynamics of a Hamiltonian system resulting from the RCSJ model. Further theoretical work in this direction is necessary to understand the full parameter space of RCSJ model in this limit.

In conclusion, driven JJs where nonlinear effects can lead to the presence of periodic attractors and chaos offers a new route to study nonlinear phenomena. The graphene junction presented in this work is a tunable platform well-suited for such studies. We hope it will be a useful toolbox to study novel phenomena at the intersection of nonlinear dynamics and condensed matter physics.

\section{Acknowledgements}
We thank Edward Ott and Gleb Finkelstein for useful conversations, and to Andrew Seredinski for assistance with sample characterization. This work is supported by Army Research Office Award W911NF-18-2-0075 (F. Y. and J. R. W.), the Physics Frontier Center at the Joint Quantum Institute (PHY-1430094) (M. T. W.), Elemental Strategy Initiative conducted by the MEXT, Japan and the CREST (JPMJCR15F3) (K. W. and T. T.), and Army Research Office Award W911NF-16-1-0132 (M. H.-R. and F. A). 

\section{Appendix A: Variation in Shapiro Diagram of the Broken Node}
\begin{figure*}
    \centering
    \includegraphics[scale=0.8]{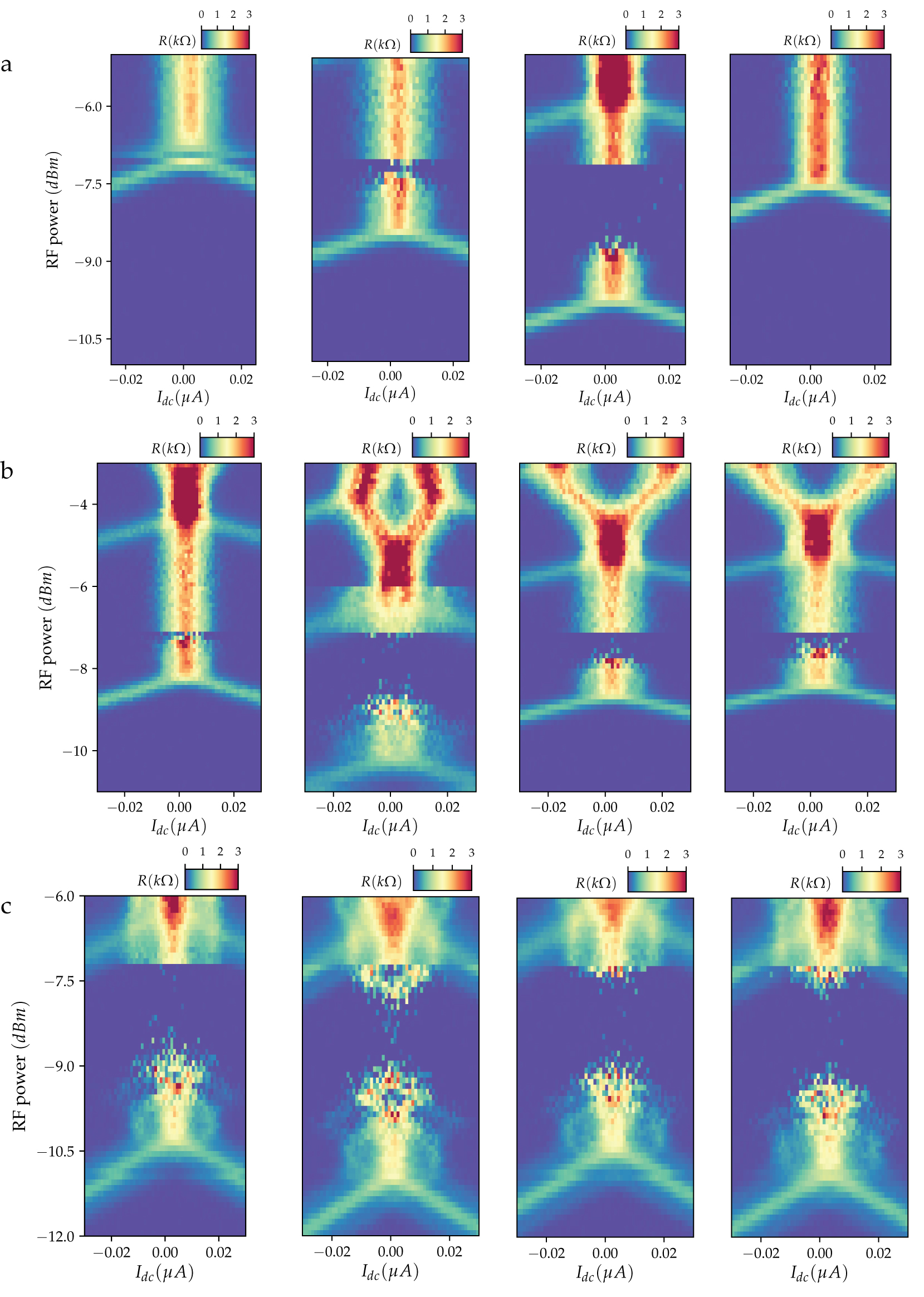}
    \caption{(a) (left to right) $f$ changes as 6.310 \si{\giga\hertz}, 6.320 \si{\giga\hertz}, 6.330 \si{\giga\hertz} and 6.370 \si{\giga\hertz} while $V_{BG}$ is fixed to CNP and no magnetic field is applied. (b) (left to right) $V_{BG}$ changes as $-0.55$ \si{\volt}, $-0.85$ \si{\volt}, $-1.05$ \si{\volt} and $-1.15$ \si{\volt} while $f$ is fixed to 6.350 \si{\giga\hertz}. (c) $B$=$0.3$ \si{\milli\tesla}, $f$ is fixed to $6.350$ \si{\giga\hertz} and we change $V_{BG}$ as $-0.45$ \si{\volt}, $-0.55$ \si{\volt}, $-0.65$ \si{\volt} and $-0.75$ \si{\volt}.}
    \label{fig-app1}
    \end{figure*}

In order to understand the origin of the broken node region, we make small changes in the external parameters that affect the junction physics. Fig.~\ref{fig-app1} shows the changes in the broken node region for small changes in $f$, $V_{BG}$ and magnetic field. In Fig.~\ref{fig-app1}(a), from left to right, $f$ increases as from 6.310 \si{\giga\hertz}, 6.320 \si{\giga\hertz}, 6.330 \si{\giga\hertz} and 6.370 \si{\giga\hertz} while $V_{BG}$ is fixed to CNP and $B$=0. In Fig.~\ref{fig-app1}(b), from left to right, $V_{BG}$ changes as $-0.55$ \si{\volt}, $-0.85$ \si{\volt}, $-1.05$ \si{\volt} and $-1.15$ \si{\volt} while $f$ is fixed to 6.350 \si{\giga\hertz}. In Fig.~\ref{fig-app1}(c), we ramp the magnetic field to $0.3$ \si{\milli\tesla}, fix $f$ to $6.350 $\si{\giga\hertz} and change $V_{BG}$ as $-0.45$ \si{\volt}, $-0.55$ \si{\volt}, $-0.65$ \si{\volt} and $-0.75$ \si{\volt}.

\section{Appendix B: Simulated Shapiro Diagram in the Overdamped Regime}\label{app:overdamped}
\begin{figure}
    \centering
    \includegraphics[width=\linewidth]{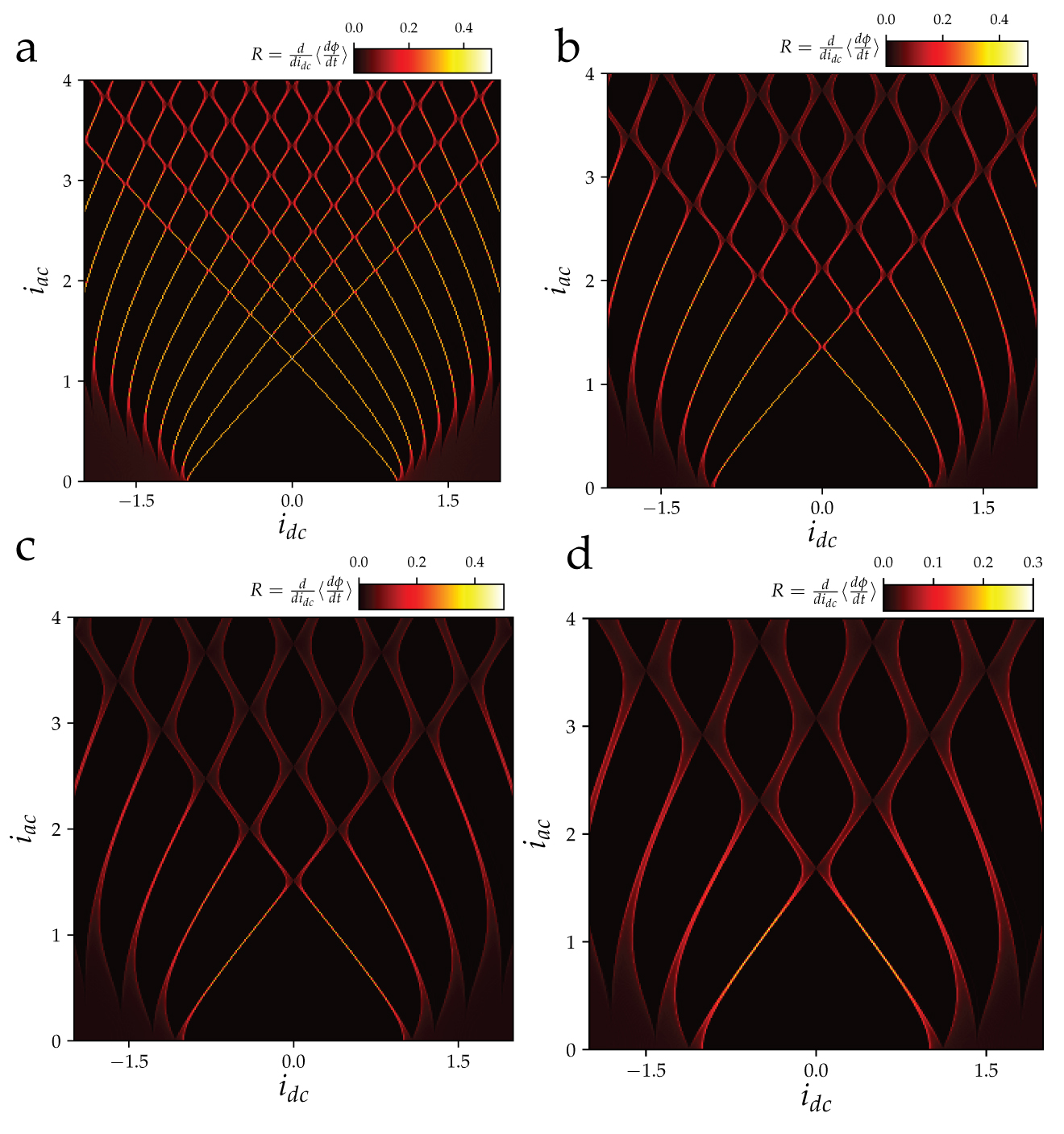}
    \caption{Simulated Shapiro diagrams in the overdamped limit From a to d, we plot the differential resistance as a function of $i_{dc}$ and $i_{ac}$ for an overdamped model with normalized frequency $\Omega$ taking on values $0.2, 0.3, 0.4$ and $0.5$ respectively. The resistive contours defining the steps meet in single points. }
    \label{fig-app2}
\end{figure}

A junction is said to be overdamped if the capacitance in the RCSJ model is small. In the limit $C = 0$, we can rewrite the evolution equation for $\phi$ as:

\begin{align}
    \frac{\hbar}{2 e R} \frac{d\phi}{dt} + I(\phi) = I_{dc} + I_{ac}\sin(\omega t)
\end{align}
We recast this equation into a dimensionless form by transforming $t \rightarrow t \frac{\hbar}{2 e R I_c}$:

\begin{align} \label{eqn-rcsj}
    \frac{d\phi}{dt} + i(\phi) = i_{dc} + i_{ac} \sin(\Omega t) 
\end{align}
where $\Omega = \omega \frac{\hbar}{2 e R I_c}$ is a normalized frequency. Note that as compared to the full RCSJ model (eqn. \ref{eqn-rcsj-model}), this system defines a flow in a two-dimensional phase space and its dynamical behaviour is constrained to take on fixed points and limit cycles by the Poincare-Bendixson theorem \cite{ott2002chaos,guckenheimer_nonlinear_2002}. In fact, it is possible to solve this equation analytically, and the Shapiro step widths take a form given by Bessel functions \cite{tinkham2004introduction}. For completeness, we reproduce Shapiro diagrams as a function of varying frequency $\Omega$ in Fig.~\ref{fig-app2}. Such overdamped diagrams have been observed before in graphene junctions \cite{heersche_bipolar_2007}.

\section{Appendix C: Underdamped Shapiro diagrams}\label{app:fine-shap}
\begin{figure*}[!hb]
    \centering
   \includegraphics{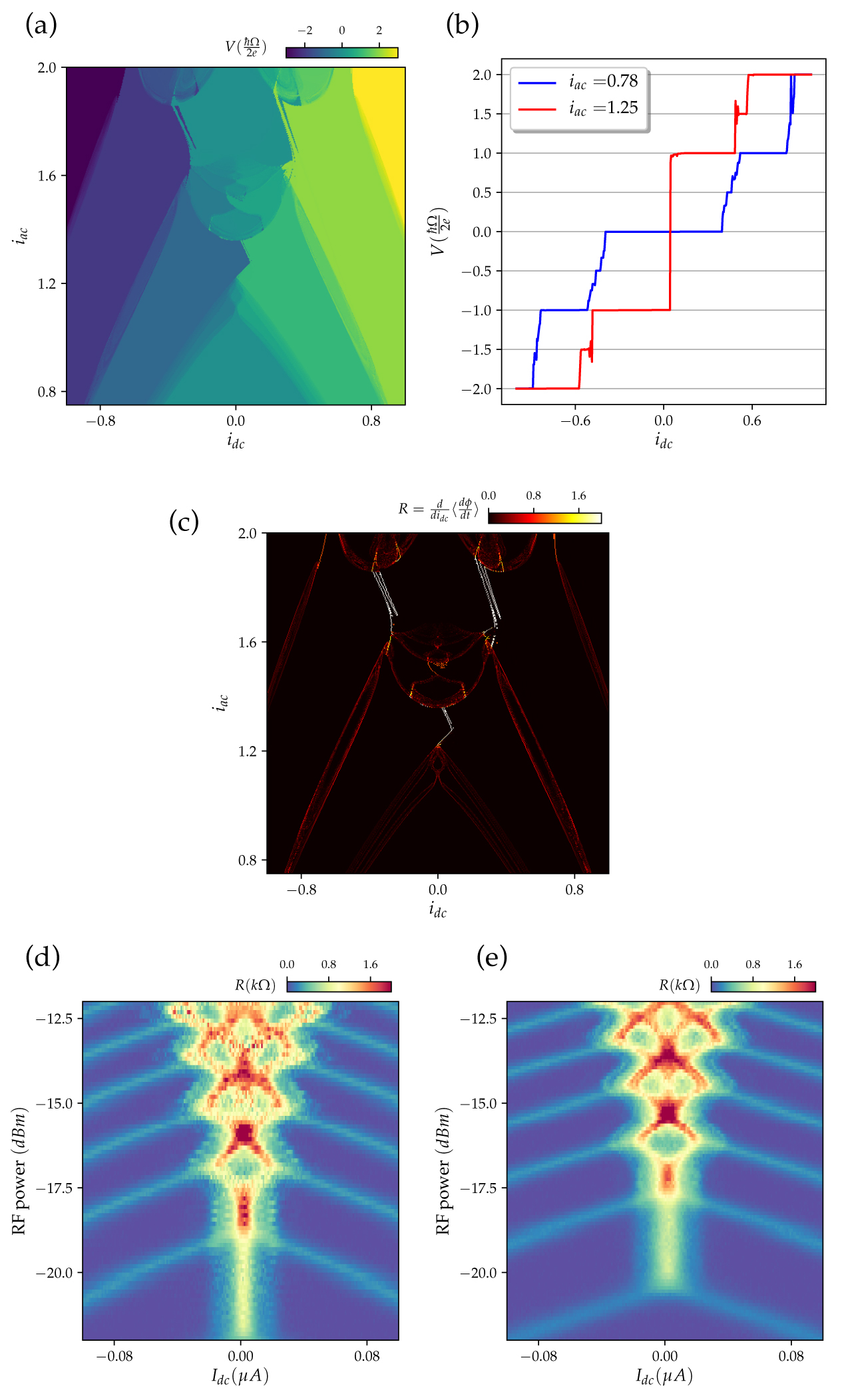}
    \caption{Appendix D: Underdamped Shapiro diagrams a. Simulated DC voltage as a function of $i_{dc}$ and $i_{ac}$ in the weakly underdamped limit for the RCSJ model, with $Q = 1.5,\Omega = 0.5$ b. Linecuts of Shapiro diagram in a at fixed $i_{ac}$ values. Both integer and fractional steps are seen. c. Simulated differential resistance Shapiro diagram for same parameters as a. d and e. Experimentally measured Shapiro diagrams at $4.000$ \si{\giga\hertz}, with no magnetic field and $V_{BG}$ set to CNP and $-0.45$ \si{\volt} respectively.}
    \label{fig-app3}
\end{figure*}

In addition to the broken node feature described in the main text, the RCSJ model in the weakly underdamped limit ($Q \sim 1$) can also be used understand the features in the resistive contours separating the Shapiro steps. Fig.~\ref{fig-app3}a and c show a simulated Shapiro diagram with DC voltage and differential resistance respectively. Fig.~\ref{fig-app3}b shows linecuts at fixed $i_{ac}$ values. In addition to integer steps, fractional steps are also seen in the transition region between two integer steps. The parameters used for this simulation were $Q = 1.5,\Omega = 0.5$, and this is a closeup version of the diagram in Fig.~\ref{fig4}a near the intersection of the $\pm 1$ step.  Fig.~\ref{fig-app3}d and e show measured Shapiro diagrams at $f$ of $4.000$ \si{\giga\hertz} and $V_{BG}$ set to CNP and $-0.45$ \si{\volt} respectively. The opening and closing up of the resistive contours  in the measurement is qualitatively similar to the fine features observed for $i_ac \in (1.2,1.6)$ in the simulation.

\section{Additional data on switching time scaling with RF power}

\begin{figure}
    \centering
    \includegraphics[width=\linewidth]{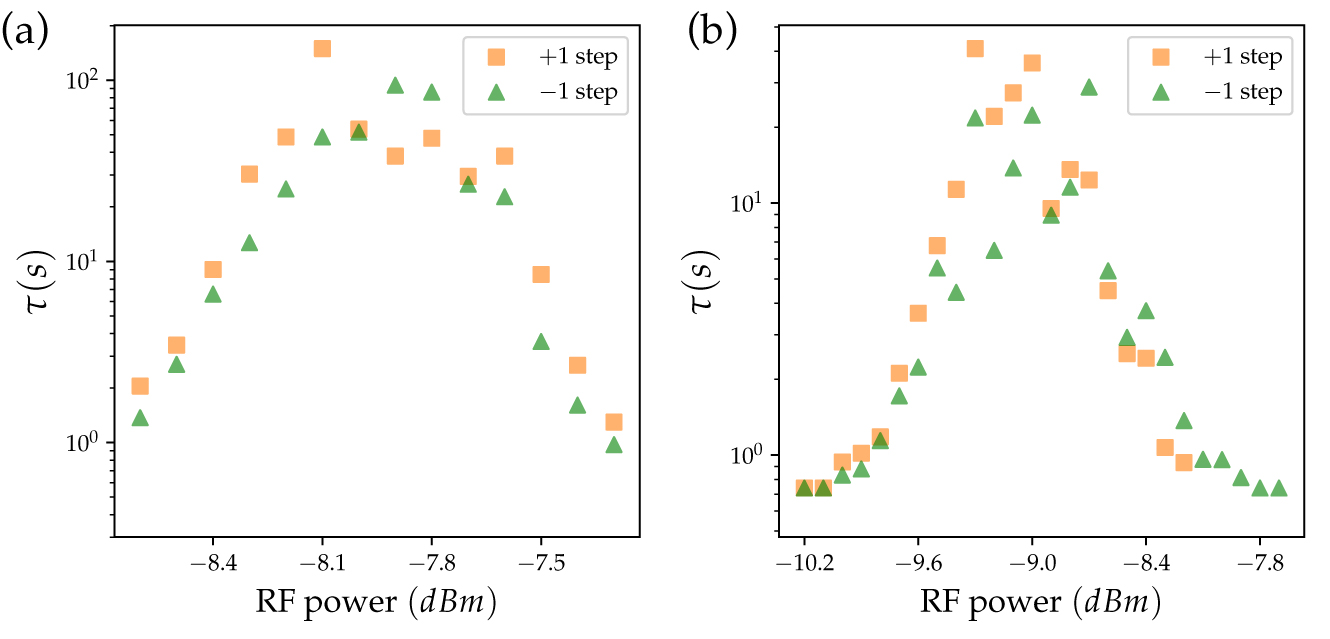}
    \caption{(a) Switching time $\tau$ vs RF power on the bistable step for $(f, B,I_{dc}) = (6.338 \si{\giga\hertz}, 0 \si{\milli\tesla}, 0 \si{\nano\ampere})$. (a) $\tau$ vs RF power on the bistable step for $(f, B,I_{dc}) = (6.35 \si{\giga\hertz},0.5 \si{\milli\tesla},1 \si{\nano\ampere})$.  }
    \label{fig-app4}
\end{figure}

The nature of the scaling relationship of switching time $\tau$ with RF power on the bistable step varies with applied frequency and the magnetic field. Fig.~\ref{fig-app4}(a) and (b) shows $\tau$ vs RF power for $(f, B,I_{dc}) = (6.338 \si{\giga\hertz}, 0 \si{\milli\tesla}, 0 \si{\nano\ampere})$ and $(f, B,I_{dc}) = (6.35 \si{\giga\hertz},0.5 \si{\milli\tesla},1 \si{\nano\ampere})$ respectively. The back gate was set to CNP.

\bibliography{ref}{}

\end{document}